\documentclass[12pt]{article}

\usepackage{amsmath}
\usepackage{amsfonts}
\usepackage{latexsym}
\usepackage{verbatim}
\usepackage{color}
\makeatletter
\newcommand{\lambdabar}{{\mathchoice
  {\smash@bar\textfont\displaystyle{0.25}{1.2}\lambda}
  {\smash@bar\textfont\textstyle{0.25}{1.2}\lambda}
  {\smash@bar\scriptfont\scriptstyle{0.25}{1.2}\lambda}
  {\smash@bar\scriptscriptfont\scriptscriptstyle{0.25}{1.2}\lambda}
}}
\newcommand{\smash@bar}[4]{%
  \smash{\rlap{\raisebox{-#3\fontdimen5#10}{$\m@th#2\mkern#4mu\mathchar'26$}}}%
}
\makeatother

\def\beq{\begin{eqnarray}}
\def\eeq{\end{eqnarray}}

\title{\bf Spin decoherence in electron storage rings ---more from a  simple
model}

\author{D.~P.~Barber$^a$ and K.~Heinemann$^b$ \thanks{mpybar@mail.desy.de,~heineman@math.unm.edu}  \\
$^a$ \small{Deutsches~Elektronen--Synchrotron, DESY, ~22607 ~Hamburg,~Germany.}
\\
$^b$ \small{Department of Mathematics and Statistics, The University of New
Mexico,} \\
\small{Albuquerque, New Mexico 87131, U.S.A. }}

\date{\today}
 
\begin{document}
\newcounter{INDEX}
\setcounter{section}{0}
\setcounter{subsection}{0}
\setcounter{equation}{0}
\renewcommand
{\theequation}{\mbox{\thesection .\arabic{equation}\alph{INDEX}}}
\maketitle

\begin{abstract}
This is an addendum to the paper ``Some models of spin coherence and
decoherence in storage rings'' by one of the authors \cite{H97}
in which spin diffusion in simple electron storage rings is studied.
In particular, we illustrate in a compact way, a key implication
in the Epilogue of \cite{H97}, namely that the exact
formalism of \cite{H97} delivers a rate of depolarisation which
can differ from that obtained by the conventional treatments of spin
diffusion which rely on the use of the derivative $\partial \hat
n/\partial\eta$ \cite{DK73,mane87,BR99}.
As a vehicle we consider a ring with a Siberian Snake and electron
polarisation in the plane of the ring (Machine II in \cite{H97} ). For
this simple setup with its one-dimensional spin motion, we avoid
having to deal directly  with the Bloch equation \cite{dbkh98,bh99}  for the polarisation density.

Our treatment, which is deliberately pedagogical,  shows that the use of $\partial \hat n/\partial\eta$
provides a very good approximation to the rate of spin depolarisation
in the model considered. But it then  shows that the exact rate of
depolarisation can be obtained by replacing $\partial \hat
n/\partial\eta$ by another derivative as suggested in the Epilogue of \cite{H97},
while giving a heuristic justification for the new derivative.

\end{abstract} 

\newpage 
\tableofcontents 
\newpage 
\setcounter{section}{0} 
\section[Introduction] {Introduction\protect \footnotemark}
\footnotetext{This paper was first prepared 
in 2000  following suggestions at that time that a Siberian Snake would help to 
preserve the polarisation in a putative high energy electron ring \cite{yale2000}.  It was then put aside
when the proponents sensibly dropped the idea.
Now, after recent work \cite{dbbq2015}, it is appropriate to make
an updated version of this paper available. }

\setcounter{equation}{0}

The paper ``Some models of spin coherence and decoherence in storage
rings'' by one of the authors \cite{H97} provides an introduction to
the use of spin-polarisation transport equations of Fokker-Planck and
Liouville type in electron storage rings. The work goes into great and careful detail 
and it uses two exactly solvable but simple model
configurations called Machine I and Machine II as vehicles.  These are perfectly aligned 
flat rings and only
the effects of synchrotron radiation and of longitudinal oscillations on electron spins lying in the machine
plane are considered. Moreover 
the orbit and spin dynamics are approximated
by smoothed equations of motion with azimuth-independent parameters
given by averages around the ring. Machine II contains a point--like
Siberian Snake in addition \cite{DK78}.
As is often the case with ``toy'' models, the exact results for the spin
distributions for Machines I and II can be expected to deliver useful insights
and expose essential features. This is indeed the case and in this paper
we use Machine II to:

\begin{itemize}
\item cover a topic mentioned, but not treated in detail in \cite{H97}, namely the 
      contrast between the exact calculations of the rate of depolarisation 
      in \cite{H97} and the conventional approach to estimating depolarisation 
      due to Derbenev, Kondratenko and Mane (DKM) \cite{DK73,mane87}. 

\item
in the process:
\begin{itemize}
\item give, and comment on, the form of the derivative $\partial \hat n/\partial\eta$
      in the DKM formula for this case,
\item show how to improve the DKM calculation in a heuristically obvious way,
      while confirming a claim in the Epilogue of  \cite{H97} and thereby calibrating the DKM
      approach in this case.

\end{itemize}
\item discuss approximations to $\partial \hat n/\partial\eta$ and 
      misunderstandings in the literature.
\item suggest further avenues of investigation.
\end{itemize}

For both model machines, polarisation build--up due to the
Sokolov--Ternov effect \cite{ST64,BR99} is ignored. In any case the
Sokolov--Ternov mechanism is ineffective in Machine II as explained in
Section 3.2.

In Section 2 we outline the models. In Section 3 we present the exact solution
for the rate of depolarisation of Machine II and the rate from the DKM theory.
Then, after comparing the two results we show how to modify the DKM calculation to get 
agreement with the exact solution. Finally, in Section 4 we make further
relevant comments and discuss other approximations and their utility.
Machine II is the simplest non--trivial exactly solvable model that we are 
aware of.

\section{The models}
\subsection{Machine I}
According to (2.5) in \cite{H97}, for Machine I the Langevin equations of 
motion  for the orbit and the spin-expectation value (the ``spin'') are
\begin{eqnarray}
&&       \left( \begin{array}{c}
                \sigma '(s)       \\
              \eta'(s)            \\
              \psi'(s)
                \end{array}
         \right)
               =
       \left( \begin{array}{ccc}
                      0 &   -\kappa   & 0         \\
  \Omega_s^2 /\kappa    &  -2 ~\alpha_s/L &  0 \\
          0           &  2\pi{\tilde{\nu}}/L  &  0
                \end{array}
         \right)   \cdot
       \left( \begin{array}{c}
                \sigma(s)         \\
                 \eta(s)          \\
                 \psi(s)
                \end{array}
         \right)
    +                    \sqrt{\omega}\cdot
       \left( \begin{array}{c}
                 0             \\
                 \zeta(s)       \\
                 0
                \end{array}
         \right)
                             \; ,
\label{eq:2.1}
\end{eqnarray}
where 
\begin{itemize}
\item $s$ denotes the distance around the ring (the ``azimuth''), 
\item $L$ is the length of the ring,
\item $\sigma$ is the distance to the centre of the bunch, 
\item $\eta$ is the fractional energy deviation, 
\item $\psi(s)$  is  the angular position of the spin in the machine plane 
      w.r.t. a horizontal direction precessing according to the T--BMT
      equation \cite{jackson} at the rate 
      $2 \pi \lbrace(g-2)/2\rbrace \gamma_0 /L$ on the design  orbit,
      where $(g-2)/2$ is the gyromagnetic anomaly and 
      where $\gamma_0 = E_0/(m_{0}{c_0}^2)$ is the Lorentz factor at the
      design orbit energy $E_0$, whereby $c_0$ is the vacuum velocity of light
      and  $m_0$ is the rest mass of the electron,
\item $\alpha_s$ is the 1--turn synchrotron damping decrement,
\item $\kappa$ is the compaction factor,
\item $\Omega_s =2\pi\cdot Q_s/L$ where $Q_s$ is the synchrotron tune for  
      undamped motion.
\item $\tilde \nu = \lbrace(g-2)/2\rbrace \gamma_0$,
\item $\omega$ is $s$-independent  and is the 
      1--turn averaged stochastic kick
      strength which is expressed in terms of an $s$-independent
      curvature  ${\bar K}_{x}$  
      $$\omega =  \frac{55}{24 \sqrt{3}} r_e  {\lambdabar}_e \gamma_{0}^{5}  {{\bar K}_{x}^3} 
        :=   \frac{55}{24 \sqrt{3}} r_e  {\lambdabar}_e \gamma_{0}^{5} \frac{1}{L} \int_{0}^{L} {ds} \:{|K_x|^3},$$
      where $r_e$ in the classical electron radius, ${\lambdabar}_e$ is the reduced Compton wavelength of the electron
      and where ${\bar K}_{x}^3$ is determined by the local curvature $K_x(s)$ of the design orbit 
      in the horizontal plane of the original ring 
      via the 1-turn average in the second equality,
      \footnote{In \cite{H97} $\omega$ was written in Gaussian units and in terms of a product of factors $C_1$ and ${C_2}$. 
      Here we work in SI units and express the  product in terms of more convenient factors.} 
      
\item $\zeta$ simulates the noise kicks due to the synchrotron radiation,
      via a Gaussian white noise process, i.e. with stochastic averages 
      $<\zeta(s_1)\cdot\zeta(s_2)> = \delta(s_1-s_2), ~~<\zeta(s)> = 0$,
      where $\delta$ denotes the Dirac delta function.
\end{itemize}
More details can be found in \cite{H97} but the content of (\ref{eq:2.1})
is clear: the motion of $\sigma$ and $\eta$ is that of a damped harmonic
oscillator subject to the noise $\sqrt{\omega}~\zeta$ in the variable $\eta$.
Moreover spin is a ``passenger'', and w.r.t. the rotating reference direction
it precesses at the rate $(2 \pi {\tilde{\nu}}/L ) \eta$. Then  the noise acting on 
$\eta$ feeds through onto the spin to cause the spin diffusion which we wish to
study. Equation (\ref{eq:2.1}) is the cleanest way to package that message.

In terms of the parameters
\begin{eqnarray}
a \equiv -\kappa \; , \;\; b \equiv \Omega_s^2/\kappa 
=(4\pi^2 Q_s^2)/(\kappa L^2)
\; , \;\; c \equiv -2  \alpha_s/L\; , \;\;
\label{eq:2.2}
\end{eqnarray}
the asymptotic ($s \rightarrow +\infty$) r.m.s widths of the distributions of 
$\sigma$ and  $\eta$  are given by
\begin{eqnarray}
\sigma_{\sigma}^2 &\equiv& \frac{\omega  a}{2bc} > 0
                                                             \; ,\qquad
\sigma_{\eta}^2 \equiv-\frac{\omega}{2c}
                   = \frac{\omega  L}{4\alpha_s}
                  =   -\frac{b}{a} \sigma_{\sigma}^2  >  0 \; .
\label{eq:2.3}
\end{eqnarray}

We also need $d \equiv 2 \pi {\tilde{\nu}} /L$.  For Machine I the 1-turn periodic solution to the T-BMT equation on the design orbit, $\hat n_0$, is perpendicular to the machine plane
and ${\tilde{\nu}}$ is the {\em design-orbit spin tune},
$\nu_0$, namely the number of spin precessions around $\hat n_0$ per
turn around the ring for a particle on the design orbit. 

For the HERA electron ring \cite{bar95b} running at about 27 GeV the values of the parameters are 
%

\begin{eqnarray}
 Q_s     &\approx& 6.0 \cdot 10^{-2}, \quad
 ~~~~~~\alpha_s \approx 3.2 \cdot 10^{-3}, \quad
 \kappa  \approx 6.9 \cdot 10^{-4}, \nonumber \\
 \omega  &\approx& 2.0 \cdot 10^{-12} \, {\rm m}^{-1}, \quad
 L       \approx 6.3 \cdot 10^{3} \, {\rm m},   \!\!    \quad
 d       \approx 6.2 \cdot 10^{-2} \,  {\rm m}^{-1}, \quad
\nonumber
\end{eqnarray}
and we adopt these parameters to illustrate our arguments. Then
%
\begin{eqnarray}
a      \approx -6.9 \cdot 10^{-4}, \quad 
b      \approx  5.2 \cdot 10^{-6} \,  {\rm m}^{-2}, \quad 
a b    \approx -3.6 \cdot 10^{-9}  \, {\rm m}^{-2}, \quad
c      &\approx& -1.0 \cdot 10^{-6}  \, {\rm m}^{-1}, \nonumber 
\end{eqnarray}
while
\begin{eqnarray}
&&\sigma_{\eta}^2 \approx 1.0\cdot 10^{-6} \; , \qquad
\sigma_{\sigma}^2  \approx  1.3 \cdot 10^{-4} \, {\rm m}^2 \; , \qquad
\label{eq:2.4}
\end{eqnarray}

For Machine I, the corresponding $\sigma_{\psi}^2$ depends on the initial distribution of
$\psi$ and an interesting point there is that the distribution of $\psi$
reaches a stationary form after a few damping times. Details can be found in
\cite{H97,BBHMR94a,BBHMR94b}. This is in contrast to
earlier calculations which suggest that the $\psi$ distribution spreads out
with a width proportional to $\sqrt {s}$ \cite{ky83}.

However, when viewed from the fixed machine coordinates the spin distribution 
is not stationary. On the contrary it is rotating continuously
in the machine plane at the rate $2 \pi {\nu_0} /L$
and cannot be periodic if ${{\nu_0}} \neq {\rm integer}$.
Such a distribution cannot be handled using DKM methods for reasons to be
explained in the Commentary.

\subsection{Machine II}
For Machine II a single point--like Siberian Snake is included at $s = 0$. 
This rotates a spin by the angle $\pi$ around the radial direction 
independently of $\sigma$ and $\eta$, so that it is a ``Type 2'' snake
\cite{mont84}.
In Machine I, $\hat n_0$, is perpendicular to the machine plane. But for Machine II
$\hat n_0$ 
is in the horizontal plane and is given by (3.3) in \cite{H97}
\footnote{For clarity we omit the subscript II used in \cite{H97}.}:
\begin{eqnarray}
 \hat {n}_{0}(s)  &\equiv&
\cos\biggl(g_6(s)\biggr)  \hat{e}_1+
\sin\biggl(g_6(s)\biggr)  \hat{e}_2 \; ,
\label{eq:2.5}
\end{eqnarray}
where $\hat{e}_1$ and $\hat{e}_2$ are unit vectors transverse and 
parallel to the design orbit respectively and
\begin{eqnarray}
 g_6(s) &\equiv& d
    \biggl(s-L/2-L {\cal G}(s/L)\biggr)  \; ,
\label{eq:2.6}
\end{eqnarray}
where the ``stairway'' function $\cal G$ is defined by:
\begin{eqnarray}
 {\cal G}(s/L) &\equiv&
   N  \qquad\qquad {\rm if} \; N < s/L < (N+1) \; ,
\label{eq:2.7}
\end{eqnarray}
for which $N$ is an integer. In the range $0  < s < L$ one has 
\begin{eqnarray}
 g_6(s) = d  \biggl(s-L/2 \biggr)  \; .
\label{eq:2.8}
\end{eqnarray} 

For Machine II it is this horizontal $\hat{n}_{0}(s)$ which is taken as the
reference direction so that for Machine II $\psi$ is the angle between a spin
and  $\hat{n}_{0}(s)$.
The conclusions of this paper are the same for
a snake with a longitudinal rotation axis (``Type 1''), provided, for example,
in the real world, by a solenoid. In that  case the spin motion on the design 
orbit is illustrated  in Figure 8 in \cite{mont84}.
The reader can easily 
make a corresponding sketch for our Type 2 snake.

Owing to the presence of the snake, $\nu_0$  is 
not ${\tilde {\nu}}$ but 1/2 \cite{H97,mont84}.
Thus the condition {\em spin--orbit resonance} occurs around $Q_s = 1/2$. Near 
resonance,  spin diffusion effects can be particularly strong.
\footnote{The resonance condition should be more
precisely expressed in terms of the {\em  amplitude-dependent spin
tune} \cite{mont98,beh2004a}. But for typical electron/positron rings the amplitude-dependent 
spin tune  differs only insignificantly from $\nu_0$.}

\section{The polarisation evolution for Machine II}
\subsection{Using exact distribution functions}
\setcounter{equation}{0}

We now look again at the exact evaluation of  the distribution functions for 
Process 3 of Machine II \cite[Section 3.5]{H97}. For this  process 
the orbital phase-space distribution is in its stationary state and 
the spins are initially all parallel to $\hat n_0$.
Then by (3.58) and (3.62) in \cite{H97}, and
after transients have died away in the first few damping times the 
polarisation of the whole beam evolves like
\begin{eqnarray}
&&  ||\vec P_{tot}^{w_3}(s)|| \propto
 \exp\biggl(-\frac{g_{14}(s)}{2}\biggr)
\cdot\exp\biggl(- \frac{s ~g_{15}}{2}\biggr) \;  \qquad
\label{eq:3.1}
\end{eqnarray}
where the function ~$g_{14}$ is 1--turn periodic and $g_{15}$, which is 
positive, is given by
\begin{eqnarray}
g_{15} &\equiv&
\frac{2 ~d^2 ~g_{11} ~\sigma_{\eta}^2}
   {~a ~b ~L ~\lambda}
    \biggl( 2 \lambda \sinh(c L/2) -  c \sin(\lambda L)  \biggr) \; , 
\label{eq:3.2}
\end{eqnarray}
with $\lambda \equiv \sqrt{-a b - c^2/4}$ and
$g_{11} \equiv {1}/\lbrace{\cosh(c L/2)+\cos(\lambda L)}\rbrace $.

Whereas  $\lambda_0 \equiv \sqrt{-a b}= \Omega_s$ is proportional to the 
synchrotron tune in the absence of damping, 
$\lambda \equiv \sqrt{-a b - c^2/4}$ is the corresponding tune in the 
presence of damping. Since $c^2 << -a b$, ~~$\lambda \approx \lambda_0$.

At large times the polarisation vanishes:
\begin{eqnarray}
      ||\vec P_{tot}^{w_3}(+\infty)|| &=& 0 \; .
\label{eq:3.3}
\end{eqnarray}
The depolarisation rate w.r.t. distance is:
\begin{eqnarray}
   \frac{1}{\tau_{spin}} &\equiv& \frac{g_{15}}{2} \; .
 \label{eq:3.4}
\end{eqnarray}
For the HERA electron ring parameters listed above one gets
\begin{eqnarray}
\tau_{spin} &\approx& 7.6\cdot 10^7\, {\rm m} \; ,
 \label{eq:3.5}
\end{eqnarray}
which corresponds to about 12000 turns, i.e. about 250 milliseconds.

For the HERA parameters, and when transients have died away after about 1000 turns
($s \approx 6.3 \cdot 10^6$ m), 
the variation of $g_{14}(s)$ over a turn causes a variation of $g_{14}(s) + s g_{15}$
in the exponent of (\ref{eq:3.1}) of about 15 percent. 
If $Q_s$ were close to 1/2 
so that $\lambda L$ were close to $\pi$, $g_{15}$ would, 
because of its factor $g_{11}$, become very large
and $\tau_{spin}$ would be very small. This is exactly what one expects
when sitting close to a spin-orbit resonance \cite{mont98}.

Of course this model {\em only} includes the effects on spin of smoothed
synchrotron motion and radiation
in the main body of the ring. No account is taken here of the detailed
dependence of spin motion on the orbital variables in a real snake 
and there is no horizontal or vertical betatron motion.

It can be seen from \cite{H97} that the asymptotic depolarisation
rate is $g_{15}/2$ for any initial distribution of spins in the machine plane.

For comparison with the results of the next section we use the relations
$\sigma_{\eta}^2 = -{\omega}/{2c}$ and  $\lambda_0 = \sqrt{-a b}$
to obtain
\begin{eqnarray}
\tau^{-1}_{spin} = 
\frac {d^2}{\lambda_0^2}  \cdot
\frac {\omega }{2 ~c ~\lambda ~L} \cdot
 \frac{1}{\lbrace{\cosh(c L/2)+\cos(\lambda L)}\rbrace} \cdot
    \biggl( 2 \lambda \sinh(c L/2) -  c \sin(\lambda L)  \biggr) \, .
\label{eq:3.6}
\end{eqnarray}

\subsection{Using the Derbenev--Kondratenko--Mane approach}
The conventional way  to calculate the rate of 
depolarisation is to use the spin diffusion term in the 
Derbenev--Kondratenko--Mane formula \cite{DK73,mane87} for the equilibrium  
polarisation.
In this approach it is assumed that at orbital equilibrium, the combined effect of 
the depolarisation and the   S--T mechanism is to cause 
the polarisation at a point in phase space to be aligned along the vector
$\hat n$ of the {\em invariant spin field} \cite{mont98,beh2004a}. This is a special solution to the T--BMT equation along the trajectory
($\sigma(s), \eta(s)$) satisfying the
periodicity condition $\hat n (\sigma,\eta; s) = \hat n (\sigma,\eta; s+L)$.
In general $\hat n$ 
is a function of all six phase-space coordinates but
in our models only $\sigma$ and $\eta$ come into play.
In the absence of radiation, and spin-orbit equilibrium, the polarisation at a point in phase space,
which we call the {\em local polarisation} is indeed 
aligned along $\hat n$ \cite{BV2006}. Then, since the characteristic times for the action
of synchrotron radiation, namely the damping time and the polarisation and depolarisation times,
are very large compared to the characteristic times for the orbital and spin-precession dynamics, the above assumption 
about the direction of the local polarisation for  electrons with radiation is reasonable since the orbital 
and spin dynamics still dominate on short time scales, at least away from spin-orbit resonances. We return to this in
Section 3.4.
The DKM approach also assumes that the value of the local polarisation and its rate of change are independent 
of the position in phase space.  These two assumptions are motivated by recognition that the stochastic   
photon emission and damping cause electrons to continually diffuse through phase space and thereby, in the end, 
have effectively interchangeable histories.

According to the DKM approach,
with a stable phase-space distribution, the local polarisation settles to the
equilibrium value
\begin{eqnarray}
           {P}_{\rm dkm} &=&
             -\frac{8}{5\sqrt{3}}
     \frac{
   { \oint {ds} \left<  {|K (s)|^{3}}
              \hat{b}
              \cdot
      (
       \hat{n}-
                 \frac{\partial{\hat{n}}}
                      {\partial{\eta}}
                           )
          \right>_{s}
                                    }
                                      }
          {
   {\oint {ds} \left< {|K (s)|^{3}}
          ( 1-
              \frac{2}{9}
     { (
       \hat{n}\cdot\hat{s}
                           )}^{2}
              +
      \frac{11}{18}
      \left(
                 \frac{\partial{\hat{n}}}
                      {\partial{\eta}}
                                        \right)^{2} \, )
          \right>_{s}
                                    }
                                     }
\label{eq:3.7}
\end{eqnarray}
where $\hat s$ = direction of the particle motion, 
$\hat b = ({\hat s} \times {\dot{\hat s}})/|{\dot{\hat s}}|$,
$K(s)$ is the curvature (in the horizontal or vertical planes) and $<\ >_{s}$ denotes an average over phase space at 
azimuth $s$. The unit vector $\hat b$ is the magnetic field direction if the 
electric field  vanishes and the motion is perpendicular to the magnetic field.
The polarisation of the beam as a whole is
\begin{eqnarray}
  {\vec  P}_{\rm dkm}(s)  \ =\ P_{\rm dkm}~ \langle \hat{n} \rangle_{s}
\label{eq:3.8}
\end{eqnarray}
In the DKM formula, the depolarisation rate w.r.t. time is  

\begin{eqnarray}
\tau^{-1}_{_{\rm dep}}  &=& \frac{5\sqrt{3}}{8}
 r_{\rm e} {\lambdabar_e} c_0 \gamma^{5} \frac{1}{L}
                 \oint ds\,
                           \left<\, {{|K(s)|^{3}}
\frac{11}{18}\, \left( {\frac {\partial{\hat{n}}} {\partial{\eta}}}\right)^{2}}
                     \right>_s 
\label{eq:3.9}
\end{eqnarray}
where, by the very nature of the DKM approach, it is assumed that 
all transients have died away. By (\ref{eq:3.8}) this is also the depolarisation rate of the whole beam. 
For a ring with constant curvature ${\bar K}_x$ (\ref{eq:3.9}) becomes
$$\tau^{-1}_{_{\rm dep}}  = \frac{55} {48 \sqrt{3}}
 r_{\rm e} {\lambdabar_e} c_0 \gamma^{5} \frac{ {\bar K}_x^3 } {L}
                 \oint ds\,
                           \left<\, \left( \frac {\partial{\hat{n}}} {\partial{\eta}}\right)^{2} \right>_s  $$
so that the depolarisation rate w.r.t. distance is 
\begin{eqnarray}
(c_0 ~\tau_{_{\rm dep}})^{-1}  &=& \frac{\omega}{2 L}
                 \int\limits^{L}_0 ds\,
\left<\, { \left( {\frac {\partial{\hat{n}}} {\partial{\eta}}}\right)^{2}}
                                                             \right>_s  \; .
\label{eq:3.10}
\end{eqnarray}

To evaluate this we need to know $\hat{n}$ at each azimuth and at each point 
in phase space. 
From (B.1) in Appendix B in \cite{H97}, in the arc the T--BMT equation reads 
as 
\begin{eqnarray}
 \frac{\partial \hat{n}}{\partial s} &=&
  -a ~\eta 
~\frac{\partial \hat{n}}{\partial \sigma}
- b~\sigma ~\frac{\partial \hat{n}}{\partial \eta}
     +  d (1 + \eta) \hat{e}_3  \times   \hat{n} \; .
\label{eq:B.1}
\end{eqnarray}
This is in fact just the partial differential form of the T--BMT equation
\begin{eqnarray}
 \frac{\partial \hat{n}}{\partial s}
+ ~\frac{d \sigma}{d s} \frac{\partial \hat{n}}{\partial \sigma}
+ ~\frac{d \eta}{d s}   \frac{\partial \hat{n}}{\partial \eta}
 &=&
     d (1 + \eta) \hat{e}_3  \times   \hat{n} \; 
\label{eq:B.1000}
\end{eqnarray}
for which the derivatives ${d \sigma} / {d s}$ and ${d \eta} / {d s}$
have been obtained from (\ref{eq:2.1}) by switching off the radiation.

By choosing the ansatz 
\footnote{For clarity we still omit the subscript II used in \cite{H97}.}
\begin{eqnarray}
 \hat{n}   &\equiv&
\cos(f) \hat{e}_1  +
\sin(f) \hat{e}_2  \; 
\label{eq:3.11}
\end{eqnarray}
(\ref{eq:B.1}) becomes
\begin{eqnarray}
&& \frac{\partial f}{\partial s} =
-a ~\eta ~
\frac{\partial f}{\partial \sigma}
- b ~\sigma ~\frac{\partial f}{\partial \eta}
          + d ~\eta  +  d  \; .
\label{eq:3.12}
\end{eqnarray}
Then by taking the snake into account and enforcing 1--turn periodicity 
we obtain
\begin{eqnarray}
&& f(\sigma,\eta;s)  =
   g_6(s)  + \sigma  g_{19}(s)  +  \eta  g_{20}(s) \; ,
\label{eq:3.13}
\end{eqnarray}
with
\begin{eqnarray}
&& g_{19}(s) =
  \frac{d ~b}{\lambda_0^2}
     \frac{1}{1+\cos(\lambda_0 L)} \biggl\lbrack
\cos\biggl(\lambda_0 \lbrack s-L-L {\cal G}(s/L)\rbrack\biggr)
\nonumber\\ && \qquad
+ \cos\biggl(\lambda_0 \lbrack s-L {\cal G}(s/L)\rbrack\biggr)
-\cos(\lambda_0 L) -1 \biggr\rbrack \; , \nonumber\\
&& g_{20}(s) =
\frac{d}{\lambda_0}
     \frac{1}{1+\cos(\lambda_0 L)} \biggl\lbrack
\sin\biggl(\lambda_0 \lbrack s-L-L {\cal G}(s/L)\rbrack\biggr)
+ \sin\biggl(\lambda_0 \lbrack s-L {\cal G}(s/L)\rbrack\biggr)
                           \biggr\rbrack \;  \nonumber\\&&
\label{eq:3.14}
\end{eqnarray}
where both are independent of $\sigma$ and $\eta$.

The resulting $\hat n$
is a solution of the T--BMT equation along the trajectory $(\sigma(s), \eta(s))$
and like $\hat n_0$ it lies in the machine plane.  
As required, it is 1--turn periodic in $s$ for all $\sigma$ and $\eta$ 
and reduces to $\hat n_0$ at $\sigma=0,\eta=0$.
If $\hat n$ had a component perpendicular to the machine plane it would, 
at simplest, be  2--turn periodic.

One also sees that  a  singularity in $\hat{n}$ occurs if the
fractional part of the orbital tune 
$Q_s= (\lambda_0\cdot L)/(2\pi)$ equals $1/2$, i.e. if one is at
a spin-orbit resonance defined in terms of $\nu_0$ (see Footnote 3). In that case $\hat{n}(\sigma,\eta;s)$ 
is not unique and a different formulation is needed. 

Obviously,
$\left({\partial{\hat{n}}}/{\partial{\eta}}\right)^{2} = g_{20}^2$
which is independent of $\sigma$ and $\eta$.  
Then to obtain the rate of depolarisation within the DKM framework we 
just need to evaluate 
\begin{eqnarray}
(c_0 ~\tau_{_{\rm dep}})^{-1}  &=& \frac{\omega}{2 L}
                 \int\limits^{L}_0 g_{20}^2(s')   ~ds'.
\label{eq:3.15}
\end{eqnarray}
In the range $0  < s < L$ we have 
\begin{eqnarray}
&& g_{20}(s)  =
    \frac{d }{\lambda_0}
     \frac{1}{1+\cos(\lambda_0 L)} \biggl\lbrack
\sin\biggl(\lambda_0 (s-L)\biggr)
+ \sin(\lambda_0 s)
                           \biggr\rbrack \; 
\label{eq:3.16}
\end{eqnarray}
so that we find
\begin{eqnarray}
(c_0 ~\tau_{_{\rm dep}})^{-1}  
&=& \frac {d^2}{{\lambda_0}^2} \cdot
\frac {\omega}{2 ~\lambda_0 ~L} \cdot
\frac{1}{\lbrace{1 + \cos(\lambda_0 L)}\rbrace} \cdot
\biggl (\lambda_0 L - \sin (\lambda_0 L) \biggr ) \, .
\label{eq:3.17}
\end{eqnarray}

This scales like $\gamma_0^7$ at fixed $\lambda_0$. 
Thus the depolarisation
due to noisy damped synchrotron motion becomes very strong at high energy.

For Machine II, the vector $\hat n$ is in the machine plane so that 
$\hat b \cdot \hat n$ is zero and the \\ 
S--T mechanism is inoperative \cite{BR99}. 
However, in this case the vector $\partial{\hat{n}}/{\partial{\eta}}$ 
acquires a vertical component. This leads to the phenomenon of
{\em kinetic polarisation} embodied in the term linear in $\partial{\hat{n}}/{\partial{\eta}}$ 
in the numerator in (\ref{eq:3.7}) \cite{spin96,mont84}. We will not pursue that here. 
The formalisms in \cite{DK73,mane87} leading to (\ref{eq:3.7}) and (\ref{eq:3.9})
are semiclassical and the concept of depolarisation does not immediately appear
in those papers, Nevertheless, the expression for
$\tau_{_{\rm dep}}^{-1}$ in (\ref{eq:3.9}) can be obtained from classical notions 
as in \cite{DK72, bhr92}. This fact will be useful in later discussions.

\subsection{Comparison}

Since $c ~L \approx -6.3 \cdot 10^{-3}$, and $\lambda \approx \lambda_0$ it is 
immediately clear that the result for $(c_0 ~\tau_{_{\rm dep}})^{-1}$
in (\ref{eq:3.17}) is very close to $\tau^{-1}_{spin}$ in (\ref{eq:3.6}) and
in fact
the relative difference between the two rates is $\approx 4 \cdot 10^{-6}$.

Thus the DKM estimate, that used is all analytical calculations of the
depolarisation rate, is very accurate and perfectly adequate for
Machine II with the parameters used here.
In fact it is expected to be very accurate in general.
But it is not exact.  This is
easy to understand: the DKM result is based on the impressive insight
that the equilibrium polarisation should be parallel to $\hat n$ at each point in phase
space but as we will see in the next section
the polarisation is not exactly parallel to $\hat n$.

If we set $Q_s$ close to 1/2 so that we are close to spin orbit
resonance \footnote{Of course one never tries to run a storage ring in
that way --- the r.f. cavity voltage would be enormous and the
smoothed equations of motion used in the model would no longer be
reasonable.}  the DKM estimate deviates significantly from the
exact result.  For example for $Q_s = 0.4998$ the relative difference is
about 65$\%$.
We expect this to be the case in general for realistic
rings too. But the depolarisation rate would then be so large that the
results would be of no practical interest.

\subsection{A modified DKM calculation}
Given the (admittedly small) differences between the two values for
the depolarisation rate we are motivated to explain them by exploiting
our exact expressions in \cite{H97} for the asymptotic spin distributions and thereby examine
the assumptions in Section 3.2 underlying the DKM approach.

Thus, as shown  by (2.120) in \cite{H97}, the unit vector 
$\vec{P}^w_{dir}(\sigma,\eta; s)$ 
describing the polarisation direction at each point in phase space does not(!)
obey the T--BMT equation. In particular, damping and diffusion effects must 
be included. Then at orbital equilibrium and after transforming
(3.33) in \cite{H97} into the machine frame we find
\begin{eqnarray}
&& \frac{\partial \vec{P}^w_{dir}}{\partial s}  =
-a~\eta ~
\frac{\partial \vec{P}^w_{dir}}{\partial \sigma}
- b ~ \sigma ~ \frac{\partial \vec{P}^w_{dir}}{\partial \eta}
+ \vec{\Omega}_{II}\times \vec{P}^w_{dir}
+ c ~\eta ~\frac{\partial \vec{P}^w_{dir}}{\partial \eta}
                                           \; .
 \label{eq:3.18}
\end{eqnarray}
It was shown in \cite{H97} that
after transients have died away in the first few damping times, 
$\vec{P}^w_{dir}(\sigma,\eta; s)$ becomes 1--turn periodic is $s$ although
the local polarisation itself then decreases smoothly to zero.
From now on we will denote the periodic $\vec{P}^w_{dir}(\sigma,\eta; s)$ 
by the unit vector $\hat p(\sigma,\eta; s)$ and we will now 
show how to obtain it
in a way paralleling the construction of $\hat n(\sigma,\eta; s)$. 
According to (\ref{eq:3.18}), in the arc $\hat p$ fulfills   
\begin{eqnarray}
&& \frac{\partial \hat p}{\partial s}  =
-a ~\eta ~
\frac{\partial \hat p}{\partial \sigma}
- b ~\sigma ~ \frac{\partial \hat p}{\partial \eta}
+ c ~\eta ~ \frac{\partial \hat p}{\partial \eta}
+ d (1 + \eta) \hat{e}_3 \times \hat p     \; .
 \label{eq:3.19}
\end{eqnarray}
Then, writing 
\begin{eqnarray}
 \hat{p}   &\equiv&
\cos(\tilde f) \hat{e}_1  +
\sin(\tilde f) \hat{e}_2  \; .
 \label{eq:3.20}
\end{eqnarray}
we have, in contrast to (\ref{eq:3.12}),
\begin{eqnarray}
&& \frac{\partial \tilde f}{\partial s} =
-a ~\eta ~
\frac{\partial \tilde f}{\partial \sigma}
- b ~\sigma ~\frac{\partial \tilde f}{\partial \eta}
+ c ~\eta ~  \frac{\partial \tilde f}{\partial \eta}
          + d ~\eta  +  d  \; .
 \label{eq:3.21}
\end{eqnarray}
So there is an extra term depending on the damping rate.
Taking into account the action of the snake, the 1--turn periodic solution 
for $\tilde f$ is 
\begin{eqnarray}
&& \tilde f (\sigma,\eta;s)  =
   g_6(s)  + \sigma  \tilde g_{19}(s)  +  \eta  \tilde g_{20}(s)
\label{eq:3.22}
\end{eqnarray}
where in the range $0  < s < L$

\begin{eqnarray}
\tilde g_{19}(s) =
- \frac{ d }{2 \lambda ~ a}
   ~\biggl\lbrack i ~ g_1( s )
               ~g_{11} ~\exp(-c ~ L/2)
            + i ~g_1( s - L )
               ~g_{11} ~\exp(c ~L/2)-2 ~\lambda\biggr\rbrack \; ,
\label{eq:3.23}
\end{eqnarray}
\begin{eqnarray}
  \tilde g_{20}(s) =
 - \frac{ d}{2 ~\lambda}
   ~\biggl\lbrack i ~g_2( s )
               ~g_{11} ~\exp(-c ~L/2)
            + i ~g_2( s - L )
               \ ~g_{11} ~\exp(c~L/2) \biggr\rbrack \; .
\label{eq:3.24}
\end{eqnarray}
with 
\begin{eqnarray}
g_1(s)  = i~\exp(c ~s/2)~\lbrack c ~\sin(\lambda ~s)
-2 ~\lambda ~\cos(\lambda~s)\rbrack \;,
\nonumber\\
g_2(s)  = 2i ~\sin(\lambda ~s) ~\exp(c ~s/2) \; ,
\label{eq:3.25}
\end{eqnarray}
so that (\ref{eq:3.20}) and  (\ref{eq:3.22}) are equivalent to (3.72) in \cite{H97}.
Thus $\hat p(\sigma,\eta; s) \neq \hat n(\sigma,\eta; s)$ owing to the 
dependence of $\hat p$ on $c$ but
for $c \rightarrow 0$ we see that 
$\tilde g_{19}(s) \rightarrow  g_{19}(s)$,
$\tilde g_{20}(s) \rightarrow  g_{20}(s)$ and $\hat p \rightarrow \hat n$
as one would expect.
Moreover, the
expression for $\tau^{-1}_{spin}$ in (\ref{eq:3.6}) then reduces to the expression
for $(c_0 ~\tau_{_{\rm dep}})^{-1}$ in (\ref{eq:3.14}).
The function $\tilde g_{20}(s)$ can be written in the form
\begin{eqnarray}
  \tilde g_{20}(s) =
 \frac{d}{\lambda}~g_{11}
   ~\biggl\lbrack ~ \sin(\lambda ~s) ~\exp(c ~(s - L)/2)
+  ~ \sin(\lambda ~(s - L)) ~\exp(c ~s/2)  \biggr\rbrack \; 
\label{eq:3.26}
\end{eqnarray}
and apart from the exponential factors containing $c$, it is reminiscent 
of $g_{20}$ in (\ref{eq:3.16}). 

We now turn to the local polarisation $||\vec P_{loc}^{w_3}(\sigma,\eta;s)||$. In the DKM 
picture it is assumed that at orbital equilibrium $||\vec P_{loc}^{w_3}(\sigma,\eta;s)||$ and its rate of change 
are independent of $(\sigma,\eta)$. 
As seen in  (3.71) in \cite{H97}, this is indeed the case in our model after transients have died away.
This emerges naturally -- there has been no need to appeal to heuristic arguments.
In particular, at orbital equilibrium the long-term $s$ dependence of $||\vec P_{tot}^{w_3}(s)||$ 
reflects the $s$ dependence of  $||\vec P_{loc}^{w_3}(\sigma,\eta;s)||$.
However the direction of the local polarisation is $\hat p$, not $\hat n$. Processes 4 and 5 in \cite{H97} 
reflect on this too.

As we are seeing, instead of relying on intuition, for 
Machine II one can find the exact result for the asymptotic depolarisation
rate by examining the development of the distribution functions.
In the general case, such an analytical treatment is not available.
But to stay close to the philosophy of using distribution functions one
can examine the properties of solutions of the Bloch equation for the polarisation density
\cite{H97, dbkh98, bh99} and then, as suggested in the 
Epilogue in  \cite{H97},  
$(\partial \hat n/\partial\eta)^2$ should be replaced by 
$(\partial \hat p/\partial\eta)^2$ in the expression for  
$(c_0 ~\tau_{_{\rm dep}})^{-1}$ of the DKM picture. 
It is then no surprise to find that
\begin{eqnarray}
(c_0 ~{\tilde \tau}_{_{\rm dep}})^{-1}  &=& \frac{\omega}{2 L}
                 \int\limits^{L}_0 \tilde g_{20}^2(s')   ~ds'.
\label{eq:3.27}
\end{eqnarray}
gives precisely the $\tau^{-1}_{spin}$ in (\ref{eq:3.6})!

This is also exactly what one expects given the physical picture in
\cite{DK72, bhr92} used to arrive at the use of 
$(\partial \hat n/\partial\eta)^2$ if the polarisation were parallel to 
$\hat n$. There, a simple geometrical construction was used to estimate the 
rate of change of the average of spin projections along the local polarisation 
direction, which was taken to be $\hat n$.
For a stored non--radiating beam in a stationary spin--orbit state, 
i.e. a non--radiating beam for which
the phase-space distribution and the polarisation distribution repeat 
from turn to turn after  a chosen starting $s$, the
polarisation is indeed parallel to $\hat n$ at each point in phase 
space \cite{mont98}. But in the presence of radiation the direction  
of the polarisation is $\hat p$ and if 
$\hat p$ had been used in \cite{DK72, bhr92} instead, then 
$(\partial \hat p/\partial\eta)^2$ would have been needed as we have 
already discovered by appealing to the Epilogue in \cite{H97}.
Thus the exact agreement between (\ref{eq:3.27}) and (\ref{eq:3.6})
supports the construction used in \cite{DK72, bhr92} once 
$\hat n$ has been  replaced by  $\hat p$. The necessity of using  $\hat p$
instead of $\hat n$ emphasises that the DKM approach is essentially 
perturbative beginning with  $\hat n$.

Of course, the ``improvement'' embodied in (\ref{eq:3.27}) is so small
that it is of no practical significance. Nevertheless our discussion 
demonstrates how our toy model can be used to expose details of spin
dynamics which  would not otherwise be accessible.

As seen in \cite{H97},
at the phase-space point where
$\sigma= \sigma_{\sigma}$ and $\eta= \sigma_{\eta}$ and for $Q_s$ well away from 1/2,
the angle between
$\hat n$ and $\hat p$ at the snake is given approximately by
\begin{eqnarray*}
&& \frac{d\cdot c\cdot\sigma_{\eta}}{2\cdot\lambda_0^2}\cdot
\frac{\lambda_0\cdot L-\sin(\lambda_0\cdot L)}
{1+\cos(\lambda_0\cdot L)} \; .
\end{eqnarray*}

For HERA parameters this is about 0.04 milliradians whereas the 
angle between $\hat n$ and $\hat n_0$ is of the order of  200 milliradians.
The angle between $\hat n$ and $\hat p$ is proportional to $c$ so that, as one might expect, the angle becomes
larger as stronger  damping causes the equation of motion for $\hat p$ to deviate
more from the T-BMT equation. 
Moreover, by comparing the denominators in $g_{11}$ and
(\ref{eq:3.14}) we see that $\hat p$ differs strongly from $\hat n$ near $Q_s = 1/2$, thereby 
confirming our suspicion in Section 3.2 that near spin-orbit resonances, care is needed with the argument 
based on time scales. This, then, is the origin of the failure of the near-resonance
DKM estimate for the depolarisation rate mentioned in Section 3.3.
Note that as seen in \cite{H97} the characteristic time for the decay of
transients remains the damping time, even close to the resonance.
In \cite{dk75} it is suggested that very close to spin-orbit resonances, extra terms containing delta functions
should be added to the DKM value (\ref{eq:3.9})  for the rate of depolarisation as a result of so-called uncorrelated resonance crossing.
These would be due to fluctuations in the rate of spin precession around $\hat n_0$ caused by 
energy fluctuations. See \cite{dbbq2015}. However, for Machine II spins precess only around the vertical in the arc, not
around $\hat n_0$ which is horizontal in the arc. So this effect will not be seen here.

\section{Commentary}
\setcounter{equation}{0}
Some further remarks are now in order ---some obvious, some less so.

\begin{itemize}
\item[(1)]
It is clear from the discussion above that $\hat n_0$ and $\hat n$ are two
different quantities which only coincide on the closed design orbit. 
Indeed, as we have just mentioned, the angle between them
is of the order of 200 milliradians for our parameters.
However, they have often been confused in the literature. 
For example $\partial \hat n/\partial\eta$ was originally written as
$\gamma (\partial \hat n/\partial\gamma)$ \cite{DK72,DK73} and that led some
to calculate  $\gamma_0 (\partial \hat n_0/\partial\gamma_0)$. 
See, for example \cite{mont84}. 
For general problems where all six phase-space coordinates must be included
this can give completely misleading results. In particular there would be
no resonant increase of the depolarisation rate when an orbital tune were 
close to $\nu_0$.
Nevertheless, if, as in our models, horizontal and vertical betatron
motion  are being neglected, it can happen that  
$(\gamma_0 \partial \hat n_0 / \partial\gamma_0)^2$ provides a useful initial 
approximation to $(\partial \hat n/\partial\eta)^2$. 
That is the case with Machine II as we now show.

Using (\ref{eq:2.5}) and (\ref{eq:2.8}) and the definition of $d$ and ${\tilde{\nu}}$ 
\begin{eqnarray}
&&  ( \gamma_0 \frac {\partial \hat n_0} {\partial \gamma_0} )^2 = 
    ( \gamma_0 \frac {\partial g_{6}   } {\partial \gamma_0} )^2 =
    d^2 ~(s - L/2)^2 =  \lbrace {\tilde {\nu}}~(2 \pi ~s/L - \pi)\rbrace ^2    \; .
\label{eq:4.1}
\end{eqnarray}
This is to be compared with 
\begin{eqnarray}
&& ( \frac {\partial \hat n} {\partial \eta} )^2 = 
   g_{20}^2 = \frac {d^2}{{\lambda_0}^2} \cdot
\frac {2} {\lbrace{1 + \cos(\lambda_0 L)}\rbrace} \cdot
{\sin}^2 \lambda_0 (s - L/2) \; . 
\label{eq:4.2}
\end{eqnarray}

The expression $\lbrace {\tilde{\nu}}~(2 \pi ~s/L - \pi)\rbrace ^2$ in (\ref{eq:4.1})
is used in \cite{spin96} and its origin is clear from the absence of the 
resonance factor $2/ {\lbrace{1 + \cos(\lambda_0 L)}}\rbrace$ 
which takes the
value $\approx 1.036$ instead of 1 for our value: $Q_s = 6.0 \cdot 10^{-2}$.
By replacing $(\partial \hat n/\partial\eta)^2$ in (\ref{eq:3.17}) by 
$(\gamma_0 \partial \hat n_0 / \partial\gamma_0)^2$ and using the result
\begin{eqnarray}
\frac{1}{ L}     \int\limits^{L}_0 
     ( \gamma_0 \frac {\partial g_{6}   } {\partial \gamma_0} )^2 ~ds'
 = \frac {d^2 ~L^2}{12} = \frac {(\pi \nu)^2}{3}.
\label{eq:4.3}
\end{eqnarray}
one obtains a value about 3$\%$ lower than from (\ref{eq:3.17}) 
so that in this case 
$(\gamma_0 \partial \hat n_0 / \partial\gamma_0)^2$ provides an adequate 
approximation. But of course it would become a bad approximation for a large
$Q_s$. Note that the approximation works for Machine II because the snake 
ensures that the spin tune, $\nu_0$, is far from $Q_s$.
In other situations one should never rely on this approximation.
The use of $(\gamma_0 \partial \hat n_0 / \partial\gamma_0)^2$ delivers 
the correct result for Machine II at 
$Q_s = 0$, i.e. when the energy is the same from turn 
to turn.

\item [(2)] The vector $\hat n$ is a $s$--periodic solution to the partial 
            differential equation (\ref{eq:B.1}). Whereas $\hat n_0$
            can be obtained as the unit real eigenvector of the 1--turn
            spin map on the closed orbit \cite{mont84},  $\hat n$ is 
            {\em not} the eigenvector of the 1--turn spin map beginning 
            at some  $\sigma, \eta$ and $s$ unless $Q_s$ is an integer. At each chosen fixed $s$, 
            each of the three components of $\hat n$ lies on a 
            component-specific closed curve in the 
            $(\sigma, \eta)$ plane corresponding to the 
            closed ellipse in the $(\sigma, \eta)$ plane mapped out by a  
            non--radiating particle of fixed amplitude. 
            But if a particular 
            $(\sigma, \eta)$ pair are chosen at some initial $s$  and the 
            corresponding $\hat n(\sigma,\eta; s)$ is transported according 
            to (\ref{eq:B.1}) for one turn, this $\hat n$ does not in general 
            return to its starting value so that it is {\em not} a ``closed
            spin solution''.

\item [(3)] Some authors still use the symbol  $\hat n$ when they actually
            mean $\hat n_0$! The tendency to create confusion seems to be
            deep rooted.
\item [(4)] The vector $\partial \hat n/\partial\eta$ (which is still often
            written as $\gamma \partial \hat n / \partial\gamma$) 
            is sometimes called the ``spin chromaticity''.
            We prefer the terms ``spin--orbit coupling function'' 
            or ``spin-field derivative''
            so that
            ``spin chromaticity'' can be reserved for the rate of change 
            of a amplitude-dependent spin tune w.r.t. a fractional energy change.
            In any case in the full theory, the DKM formula (\ref{eq:3.7}) 
            must be modified to include (usually) relatively small terms 
            involving derivatives of $\hat n$ w.r.t. the 
            two transverse canonical momenta \cite{mont98,bm88}
            and for such terms the name ``chromaticity'' is clearly 
            unsuitable.

\item [(5)] If $|\sigma  g_{19}  +  \eta  g_{20}| << 1$, i.e. if the angle 
            between $\hat n$ and $\hat n_0$ is small, then $\hat n$ in
            (\ref{eq:3.11}) and (\ref{eq:3.13})  can be approximated by
\begin{eqnarray}
\hat n = \hat n_0 + \biggl(\sigma  g_{19}(s) + \eta  g_{20}(s)\biggr) ~\hat m
\label{eq:4.4}
\end{eqnarray}
where 
\begin{eqnarray}
\hat m = -\sin\biggl(g_6(s)\biggr)\hat{e}_1 +
                                          \cos\biggl(g_6(s)\biggr)  \hat{e}_2 
\label{eq:4.5}
\end{eqnarray}
            is a unit vector in the
            machine plane perpendicular to $\hat n$.
            This is the ``SLIM approximation'' \cite{chao81} whereby $\hat n$
            deviates from
            $\hat n_0$ by a function linear in the phase-space 
            coordinates \cite{BR99}.
            In this approximation we again find  
            $(\partial \hat n/\partial\eta)^2 = g_{20}^2$ so that for 
            Machine II the SLIM approximation actually delivers the correct 
            result for $(\partial \hat n/\partial\eta)^2$.
            Of course, a direct application of the eigenvectors and matrices
            of the SLIM formalism delivers (\ref{eq:4.4}) too. 
\item [(6)] The reason that the DKM approach cannot be used on the spin  
            distribution of Machine I is as follows. The DKM formalism 
            assumes that transients have died away and that the polarisation is
            locally parallel to $\hat n$.
            But in Machine I the magnetic field is perpendicular to the
            machine plane everywhere so that $\hat n_0$ and $\hat n$ are also  
            perpendicular to the machine plane. Then since the spin 
            distribution of Machine I is set up in the machine plane 
            the spins are perpendicular to $\hat n$ and  the distribution 
            precesses at the rate $\nu_0$ around $\hat n_0$.
            In Machine II there is a magnetic field in the machine plane at 
            the snake and $\hat n_0$ and $\hat n$ are in the machine plane
            together with the spins themselves. In this case the polarisation
            can settle down to be almost parallel to $\hat n$.
\item [(7)]
    We have seen how the fields ${\hat n}(\sigma, \eta ; s)$ and
  ${\hat p}(\sigma, \eta ; s)$ for Machine II are found by enforcing
  1-turn periodicity while solving the partial differential equations
  (PDE) (\ref{eq:B.1}) and (\ref{eq:3.19}) respectively.  For real
  rings, the most reliable method for finding $\hat n$ is stroboscopic
  averaging \cite{hh96} for a point in phase space.  That involves
  integrating the T-BMT equation for spins along a particle trajectory through
  that point in phase space and averaging the spins.  The 1-turn
  periodicity emerges automatically from the algorithm.  For
  (\ref{eq:B.1}) the trajectory is, of course, defined by the
  relations $d \sigma / ds = a \eta$ and $d \eta /ds = b \sigma$. But
  these just define the characteristic curves of the PDE, familiar from
  the method of characteristics for solving linear PDE's \cite{ch}.
  Stroboscopic averaging for $\hat n$ relies on the fact that the
  T-BMT equation (equivalently (\ref{eq:B.1})) is linear.  The equation
  of motion for $\hat p$ in (\ref{eq:3.19}) and for real rings is also
  linear.  Thus, since stroboscopic averaging works well for $\hat n$,
  it is tempting to use it for $\hat p$.  In that case one would again
  integrate along a characteristic curve.  However, the defining
  equations for a characteristic curve would now include damping. Then
  care would be needed.  In any case stroboscopic averaging is only
  useful if the average is normalisable.
\item [(8)] In principle the asymptotic direction of the local
polarisation at points in phase space and its dependence on the distance to spin-orbit resonance could be discovered
with a Monte-Carlo spin-orbit tracking simulation \cite{db2004,dbbq2015}. This would require a large
number of particles and 
corresponding  computing power in order to sufficiently populate a sufficient number of points in phase space.
            In addition a precise 
            numerical determination of $\hat n$ at each point in phase space would be needed for comparison
            and that would require stroboscopic averaging \cite{hh96} or the SODOM algorithm \cite{ky99}. Away 
            from spin-orbit resonances the 
            required number of particles  could 
            perhaps be reduced
            by just looking (say) turn-by-turn at the projection of a spin on the plane perpendicular to
            $\hat n$ for each particle and then finding the average projection. This would
            represent the deviation of the direction of the local polarisation from $\hat n$, averaged over           
            phase space.  
 \item [(9)] Although it would be satisfying to calculate the rate of depolarisation for 
             real  electron/positron  machines by integrating  the 
            Bloch equation for the polarisation density, that would be cumbersome in practice. 
            In any case the DKM approach, as well as being elegant, provides 
            an extremely good approximation perfectly 
            suited to the  operating regimes of storage rings in operation up to now
            and it provides the only practical analytical approach 
            for real rings with arbitrary discrete magnet structures.
            Computer codes which evaluate the DKM formula numerically for real rings
            are listed in \cite{BR99}. Unfortunately, beyond the
            first order approximation of SLIM, they all require large amounts of
            computing time. 

            Nevertheless, the Bloch equation for the polarisation density, augmented by a
            Baier-Katkov-Strakhovenko expression for the influence of
            the Sokolov-Ternov effect \cite{dbkh98,bh99,beh2016}
            remains the key to a general description of polarisation
            dynamics in electron/positron rings.  Note that as
            explained in \cite{dbkh98,bh99} and hinted at in the
            Epilogue in \cite{H97}, with this combination it will be
            possible to arrive at the DKM estimates from first principles.
            A first approach, based on a semiclassical calculation,
            can be found in \cite{dk75}.  This programme will involve
            approximations but will enable analytical exploration of
            the limitations of the DKM formula without recourse to
            heuristics. See \cite{dbbq2015} for Monte-Carlo
            simulations of the depolarisation process at very high
            energy and for a study of whether the extra terms for the
            depolarisation rate mentioned in Section 3.4 are needed.

\end{itemize}

\section*{Conclusion}
\addcontentsline{toc}{section}{Conclusion} 

Analytical calculations for
the polarisation in electron/positron storage rings are usually based
on the DKM formalism. This, in turn, is based on some implicit
reasonable considerations of the various time scales involved in
spin-orbit motion. See, for example, Figure 1 in \cite{mont84}.
Nevertheless there are open questions about the applicability of the
DKM formula at very high energy \cite{dbbq2015}.

It therefore seems desirable to return to first principles and obtain
the rate of depolarisation from a study of the solutions of the Bloch
equation for the polarisation density \cite{dbkh98,bh99}. To obtain
the equilibrium polarisation, the Sokolov-Ternov effect must be included 
via the Baier-Katkov-Strakhovenko
formalism.

This paper has used a simple, non-trivial model of spin diffusion to
show that the DKM expression for the depolarisation rate is 
only a good approximation in this case.  Although the Bloch equation
for the polarisation density was not used directly, its use was
implicit and our results support the claim in the Epilogue of
\cite{H97} that the derivative $\partial \hat n/\partial\eta$ should
be replaced by $\partial \hat p/\partial\eta$.  Nevertheless, further
work will be needed to properly establish the range of applicability
of the basic DKM formalism and to address the open questions 
mentioned in \cite{dbbq2015}.

\section*{Acknowledgments}
\addcontentsline{toc}{section}{Acknowledgments}
We thank Zhe Duan for helpful questions and discussions.

\end{document}